\PassOptionsToPackage{utf8}{inputenc}
\documentclass{bioinfo}
\copyrightyear{2020} \pubyear{2020}

\access{Advance Access Publication Date: Day Month Year}
\appnotes{Manuscript Category}

\usepackage{xspace}
\usepackage{amsmath}
\usepackage{subcaption}
\usepackage{relsize}

\newcommand{\model}[0]{\texttt{DTI-GAT}\xspace}

\begin{document}
\firstpage{1}

\subtitle{Subject Section}

\title[DTI-GAT]{Drug-Target Interaction Prediction with Graph Attention networks}
\author[Wang \textit{et~al}.]{Haiyang Wang\,$^{\text{\sfb $\dagger$,1}}$, Guangyu Zhou\,$^{\text{\sfb $\dagger$,2}}$, Siqi Liu\,$^{\text{\sfb 2}}$, Jyun-Yu Jiang\,$^{\text{\sfb 2}}$ and Wei Wang\,$^{\text{\sfb 2,}*}$}

\address{$^{\text{\sf 1}}$Zhiyuan College, Shanghai Jiao Tong University, Shanghai, 200240, China and \\
$^{\text{\sf 2}}$Department of Computer Science, University of California, Los Angeles, 90095, USA}

\corresp{$^\ast$To whom correspondence should be addressed.  ${\dagger}$ These authors contributed equally to this work.}



\abstract{\textbf{Motivation:} Predicting Drug-Target Interaction (DTI) is a well-studied topic in bioinformatics due to its relevance in the fields of proteomics and pharmaceutical research. Although many machine learning methods have been successfully applied in this task, few of them aim at leveraging the inherent heterogeneous graph structure in the DTI network to address the challenge. 
For better learning and interpreting the DTI topological structure and the similarity, it is desirable to have methods specifically for predicting interactions from the graph structure.\\
\textbf{Results:} 
We present an end-to-end framework, \model
(\textbf{D}rug-\textbf{T}arget \textbf{I}nteraction prediction with \textbf{G}raph \textbf{AT}tention networks) for DTI predictions. \model incorporates a deep neural network architecture that operates on graph-structured data with the attention mechanism, which leverages both the interaction patterns and the features of drug and protein sequences. \model facilitates the interpretation of the DTI topological structure by assigning different attention weight to each node with the self-attention mechanism. Experimental evaluations show that \model outperforms various state-of-the-art systems on the binary DTI prediction problem. Moreover, the independent study results further demonstrate that our model can be generalized better than other conventional methods. 
\\
\textbf{Availability:} The source code and all datasets are available at https://github.com/Haiyang-W/DTI-GRAPH\\
\textbf{Contact:} \href{wanghaiyang@stu.pku.edu.edu}{wanghaiyang@stu.pku.edu.edu}, \href{weiwang@cs.ucla.edu}{weiwang@cs.ucla.edu}\\
}

\maketitle

\section{Introduction}\label{sec:intro}

Detecting drug–target interactions (DTIs) potentially facilitates therapeutic target identification \citep{xia2010semi,petta2016modulation} and novel drug design \citep{skrabanek2008computational, ay2007drug, janga2009structure, kuhn2008large}. 
Until quite recently, pharmacological effects were often discovered using primitive trial and error procedures, such as applying plant extracts on living systems and observing the outcomes \citep{singh2016biological}. 
However, experiment-based methods remain expensive, labor-intensive and time-consuming \citep{dickson2004key, kola2004can, kapetanovic2008computer}. 
Evidently, there is an immense need for reliable computational approaches to identify and characterize DTIs, hoping to accelerate the pace and reduce the cost of drug development. 
%

With the rapid development of machine learning techniques, various computational prediction approaches have been proposed to predict drug-target interactions. 
\cite{yamanishi2010drug} integrated the relationship among the pharmacological space, the chemical space, and the topology of drug–target interaction networks to predict the associations between drugs and targets, and their experimental results have demonstrated that drug–target interactions are more correlated with pharmacological effect similarity than with chemical structure similarity. 
According to the similarity of chemical information, \cite{keiser2009predicting} proposed a method to explore the associations between drugs and targets. They selected 30 of predicted results for biological experiments and finally confirmed 23 with interrelationships. 
\cite{wang2010computationally} used supervised machine learning methods to predict the relationship between drugs and targets. To solve the problem of sample imbalance, they are collecting the positive samples from the database, and the negative samples using the random selection method. The input features of the classifier consist of the chemical structure of the drug and the sequence information of the protein. 
\cite{chen2012drug} developed a novel method of Network-based Random Walk with Restart on the Heterogeneous (NRWRH) network to predict potential drug–target interactions on a large scale. The excellent experimental results show that the proposed method can discover new potential drug–target interactions for drug development.
These approaches provide feasible solutions to the problem. However, the extracted features used in these approaches only have limited coverage on interaction information, since they are dedicated to specific facets of the protein and the drug profiles.

To alleviate the inadequacy of statistical learning methods, deep learning algorithms provide the powerful functionality to represent large-scale raw data for different tasks and thus facilitate the learning of latent patterns in the data \citep{lecun2015deep}.
Recently, deep learning architectures have produced powerful systems to address several estimation problems related to protein sequences,
such as protein-protein interaction \citep{chen2019multifaceted}, protein binding affinity upon mutation \citep{zhou2020mutation}, and protein structural changes \citep{senior2020improved} estimation. 
These works typically use convolutional neural networks (CNNs) for automatically selecting local features, recurrent neural networks (RNN) that aim at preserving the contextualized and long-term ordering information or the combination of both CNNs and RNNs.
In contrast, fewer efforts have been made to capture the pairwise interactions of the protein drug interaction with deep learning,
which remains a non-trivial problem with the following challenges:
(i) Characterization of the proteins and drugs requires a model to effectively filter and aggregate their local features, while preserving significant contextualized and sequential information of the amino acids and drug fingerprints;
(ii) Constructing a deep neural architecture without biological insights often suffers from the interpretation issue;
(iii) 
An effective mechanism is also needed to apprehend the mutual influence of protein-drug pairs in DTI prediction. Moreover,the framework needs to be scalable to large data.
Corresponding methods, including DeepDTI \citep{wen2017deep}, RFDT \citep{wang2018computational}, DeepDTA \citep{ozturk2018deepdta} and DeepConv-DTI \citep{lee2019deepconv}, employed deep neural networks on DTI prediction tasks.
Specifically, DeepDTI built by \cite{wen2017deep} used the deep belief network (DBN), with features such as the composition of amino acids, dipeptides, and tripeptides for proteins and fingerprints for drugs. RFDT by \cite{wang2018computational} employed stacked Auto-Encoder (AE) to abstract original features into a latent representation with a small dimension. With latent representation, they trained a Random Forest (RF), which performed better than previous methods. 
DeepDTA \citep{ozturk2018deepdta} and DeepConv-DTI \citep{lee2019deepconv} both used CNN to extract local residue patterns to predict the binding affinity between drugs and targets. They performed convolution on various lengths of the subsequences' amino acids to capture local residue patterns of generalized protein classes.

The previous DTI prediction methods can be mainly separated into two categories: similarity-based methods and feature-based methods. Similarity-based methods assume that similar drugs or proteins may have similar interaction patterns. These methods use many different similarity measures based on fingerprint, chemical structure, sequence data and so on to identify drug-target interaction. Feature-based methods solve drug–target interaction prediction as a binary classification problem, such as  DeepConv-DTI \citep{lee2019deepconv}, use known drug-target pairs as positive sample. To combine both the similarity and feature vectors, the neural network models require a mechanism to represent both the interaction patterns and the features of drug and protein sequences. 

In this paper, we introduce \model
(\textbf{D}rug-\textbf{T}arget \textbf{I}nteraction prediction with \textbf{G}raph \textbf{AT}tention networks), a deep neural network architecture that operates on graph-structured data with attention mechanism. 
First, it converts the position-specific scoring matrix (PSSM) and the drug fingerprint as feature vectors for all the target proteins and drugs. 
Secondly, a DTI graph is constructed based on the similarity of feature vectors. Each node represents a protein or a drug and the nodes are lined by edges representing interactions between protein-drug, protein-protein and drug-drug separately. 
The graph attention network is then applied on the built graph to generate embeddings for each protein and drug, followed by a final decoder architecture to predict the interaction result. 
Notably, based on the assumption that different neighbors' importance are different and the DTI graph is heterogeneous, we don't know which neighbor or domain knowledge (protein or drug) is more important for a given node in the DTI prediction task. The Graph attention (GAT) layer applies self-attention mechanism, which allows for assigning different attentional weight to neighbor nodes, enabling a leap in model capacity. Furthermore, analyzing the learned weights with attention can lead to benefits in interpreting the DTI topological structure and similarity.

Our contributions are 3-fold. 
First, we provide an approach to transform the feature representations of proteins and drugs into a protein-drug interaction graph. We emphasize the need to extend the bipartite graph to a heterogeneous graph, by adding drug-drug and protein-protein similarities. 
Second, we demonstrate that the attention mechanism can automatically extract the important high-level relationships among proteins and drugs by assigning different weights to each edge and drop all structural information.
Third, we provide an extensive analysis to show the better interpretability of learned attention weights for representing the topological structure and similarity.
Last, \model is highly efficient, as the operation of the self-attentional layer can be parallelized across all edges, and the computation of output features can be parallelized across all nodes. 
\model significantly outperforms various state-of-the-art approaches on the DTI binary prediction task, which confirms the effectiveness of the graph attention strategy in identifying drug-target interactions. 

\section{Methods}\label{sec:method}
\subsection{Preliminary}

\subsubsection{Protein feature representation}
In this work, protein sequences are represented as pseudo-position specific scoring matrix (PsePSSM) features to encode the evolution and sequential information for proteins with different lengths of sequences.
Note that this setting is consistent with previous studies~\citep{shi2019predicting}.

For a target protein sequence $p_m$ with $L$ amino acid residues, we use
the position-specific scoring matrix (PSSM) as its descriptor introduced by \cite{jones1999protein}. 
The PSSM with a dimension of L $\times$ 20 can be expressed as:
\begin{equation*}
PSSM(p_m) = 
\begin{pmatrix}
E_{1\rightarrow{1}} & E_{1\rightarrow{2}} & \cdots & E_{1\rightarrow{20}} \\
E_{2\rightarrow{1}} & E_{2\rightarrow{2}} & \cdots & E_{2\rightarrow{20}} \\
\vdots  & \vdots  & \ddots & \vdots  \\
E_{L\rightarrow{1}} & E_{L\rightarrow{2}} & \cdots & E_{L\rightarrow{20}}
\end{pmatrix},
\end{equation*}
where each element $E_{i\rightarrow{j}}$ in the PSSM matrix is then normalized to the interval (0, 1) as:
\begin{equation}
    \overline{E}_{i\rightarrow{j}} = \frac{1}{1+exp(E_{i\rightarrow{j}})}
\end{equation}

To make the PSSM descriptor a uniform representation despite proteins with different lengths correspond to different numbers of rows, we represent the uniformed PSSM of $p_m$ as:
\begin{equation}
    \overline{PSSM}(p_m) = [\overline{E}_{1}, \overline{E}_{2}, ..., \overline{E}_{20}]^T,
    \label{eq:uni_pssm}
\end{equation} where $T$ is the transpose operator. 
Here, $\overline{E}_{j} = \frac{1}{L} \sum_{i = 1}^{L} \overline{E}_{i\rightarrow{j}}$ computes the average score of the residue in protein $p_m$ during the process of evolution, which is the mutation from amino acid type $i$ to $j$.
To retain the sequence information after Eq.~\ref{eq:uni_pssm}, the pseudo-position specific scoring matrix (PsePSSM) for protein $p_m$ is computed as:
\begin{equation}
    P_{\text{Pse}(p_m)}^{\lambda} = [\overline{E}_1, ..., \overline{E}_{20}, G_1^1, ..., G_{20}^1, ..., G_1^\lambda, ... G_{20}^\lambda]^T,
    \label{eq:psepssm}
\end{equation}
where 
\begin{equation}
    G_{j}^{\lambda} = \frac{1}{L - \lambda} \sum_{i = 1}^{L - \lambda}[\overline{E}_{i \rightarrow{j}} - \overline{E}_{(i + \lambda)\rightarrow{j}} ]^2, 
    (j=1,...,20;0 \leq \lambda \leq L).
\end{equation}
Here, $G_{j}^{\lambda}$ is the correlation factor of $j^{th}$ amino acid and $L$ is the continues distance along the protein sequence. Therefore, a protein sequence can be expressed as Eq.~\ref{eq:psepssm} using PsePSSM and generates a $20 + 20 \times \lambda$ dimensional feature vector.

\subsubsection{Drug feature representation}
It has been indicated in some studies that descriptors as molecular structure fingerprints can effectively represent the drug \citep{ding2017identification,yamanishi2011extracting}.
For drug molecules, we use the chemical structure of the molecular substructure fingerprints from the PubChem database \citep{kim2019pubchem}. For each drug molecule, it defines an 881- dimensional binary vector $Q$
to represent the molecular substructure, where the corresponding bits of the vector are encoded as 1s for existence of substructures and 0s for absence. 
Therefore, given a drug $d_n$, its fingerprint feature is calculated as $Q(d_n) = [q1(d_n), q2(d_n), ... q881(d_n)]$
Fingerprints property is ``PUBCHEM\_CACTVS\_SUBSKEYS'' in PubChem SDF files and is Base64 encoded, which provides a textual description by the binary data.
The drug feature representations in this work are also consistent with the previous study~\citep{wang2018computational}.

\subsubsection{Problem statement}
Given a set of $N$ proteins $P = (p_1, p_2, ..., p_N)$ and a set of $M$ drugs $D = (d_1, d_2, ..., d_M)$, our goal is to predict the interaction ($I$) between $p_m$ and $d_n$ based on the the protein's PsePSSM: $P_{\text{Pse}(p_m)}^{\lambda}$ and the drug's fingerprint: $Q(d_n)$, where $p_m \in P$ and $d_n \in D$. 
\begin{equation}
    I(p_m, d_n) = 
\left\{
            \begin{array}{lr}
            1, \quad interaction &  \\
            0, \quad no\quad interaction &  \\
            \end{array}
\right.
\label{eq:interaction_define}
\end{equation}




\subsection{\model}
\begin{figure*}[!t]
    \centering
    \includegraphics[width=\linewidth]{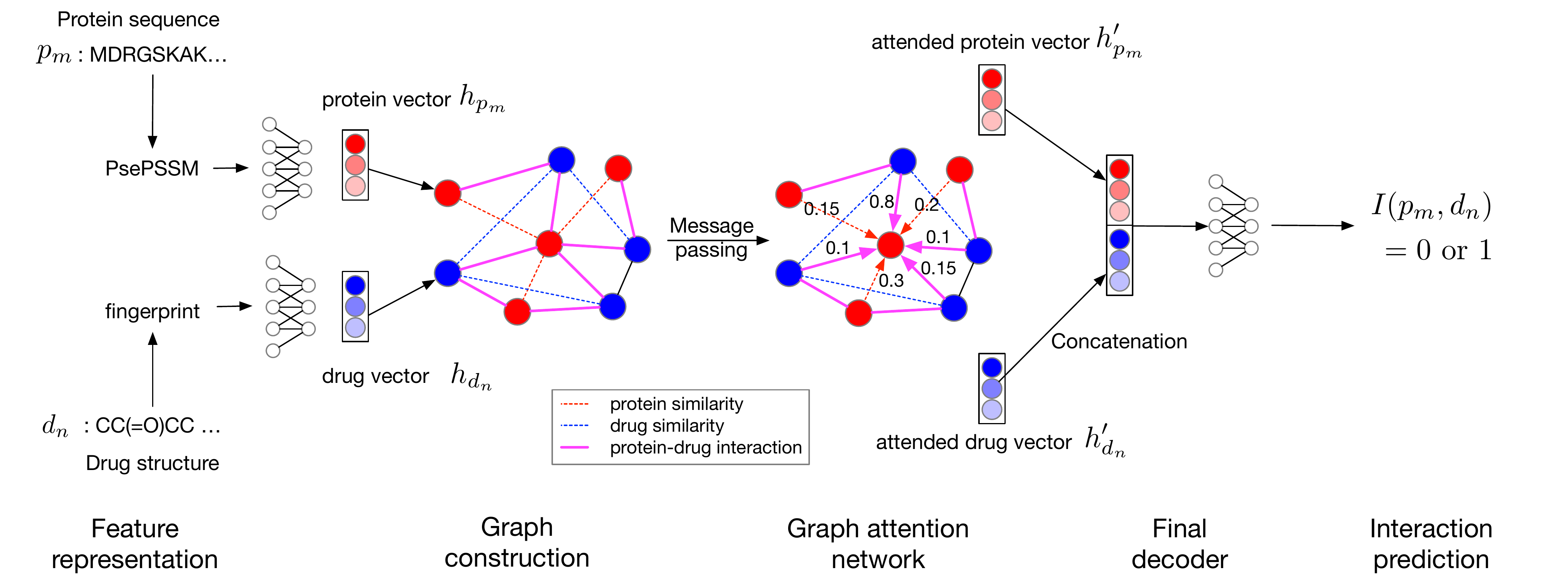}
    
    \caption{Architecture of \model.}\label{fig:dpigat}
\end{figure*}

We introduce \model, a deep neural network architecture that operates on graph-structured data with attention mechanism for the feature-based DTI prediction task. 
The overall learning architecture is illustrated in Figure~\ref{fig:dpigat}.

\model consists of the following main components: 
First, given the pseudo-position specific scoring matrix (PsePSSM) of each protein and the fingerprint of each drug, a DTI graph is constructed based on the information about interaction target and similarity of local structure. 
Then, the graph attention network is applied to the built graph to integrate higher-level information and generate embeddings for each protein and drug. 
A final decoder architecture takes a pair of protein-drug embeddings and predicts the final interaction probability.

\subsubsection{Graph construction} \label{section:graph construction}
Given a protein set $P = (p_1, p_2, ..., p_N), p_i \in R^p$, drug set $D = (d_1, d_2, ..., d_M), d_i \in R^d$, and interaction targets set $T = (t_1, t_2, ..., t_k), t_i \in \{1,-1\}$, where $p$ is the PsePSSM feature dimension, $d$ is the fingerprint dimension, we construct a DTI-Graph  $G(V,E)$.

First, we transform the PsePSSM and the fingerprint to the same dimension, so that nodes can be easily aggregated. In our experiment, we use two weight matrices, $W_p \in R^{p \times F}$ and $W_d \in R^{d \times F}$ for transforming protein and drug separately,  where F is the feature dimension, and $ReLU$ nonlinearity to transform the feature dimension.
\begin{equation}
    p^{'}_i = ReLU(W_p \cdot p_i),  \ p^{'}_i \in R^F
    \label{eq:protein feature transform}
\end{equation}
\begin{equation}
    d^{'}_i = ReLU(W_d \cdot d_i),  \ d^{'}_i \in R^F
    \label{eq:drug feature transform}
\end{equation}
Once obtained, then we combine the new protein set $P^{'} = (p^{'}_1, p^{'}_2, ..., p^{'}_N), p^{'}_i \in R^F$ and the new drug set $D = (d^{'}_1, d^{'}_2, ..., d^{'}_M), d^{'}_i \in R^F$ to the graph node set $H = (h_1, h_2, ..., h_K), h_i \in R^F$, where each node is represented as a vector. These nodes are linked by undirected edges, $e \in \{1,-1\}$, which are generated from the drug-protein interaction, drug-drug and protein-protein similarity.

The edge set, $E$, contains two different components: \textbf{interaction edge} and \textbf{similarity edge}. The interaction edge, $e_i$, is either $1$ or $-1$, referring positive relation and negative relation of two nodes. Since the positive relation is from the ground truth interaction targets set, $T$, the initial graph of the DTI is very sparse and has an imbalanced issue. To solve the above issues, the negative samples are selected randomly from the unidentified drug–target pairs. We assume that the positive sample is a small percentage of all possible samples, so there is a low probability that real interaction will be selected as a negative sample. Actually, the proportions of positive samples detected in each dataset are $0.99\%$(Enzymes), $3.49\%$(Ion channels), $2.99\%$(GPCR) and $6.40\%$(Nuclear receptors). In the experiment, we choose a negative sample with the same number of positive samples. However, the initial interaction edges are only between the drug-protein pairs, which construct a bipartite graph that limit the information flow. 
In order to aggregate more information, we transform the bipartite graph to a heterogeneous graph, by adding drug-drug and protein-protein similarities. Similarity edge, $e_s$, is based on the DTI bipartite graph and its common neighbor information. If the number of two nodes' common positive neighbor or negative neighbor is greater than a threshold, the two nodes will be connected by $1$, which means they are similar. 
\begin{equation}
    e_s(i,j) = 
\left\{
            \begin{array}{lr}
            1, \quad if \ com\_n(h_i, h_j) \textgreater \theta &  \\
            0, \quad if \ com\_n(h_i, h_j) \leq \theta &  \\
            \end{array}
\right.
\label{eq:similarity}
\end{equation}
where $com\_n$ is the function to compute the common neighbor of two nodes and $\theta$ is the threshold. 
In our experiment, we use adjacent matrix $A \in R^{K \times K}$ to represent $E$, where $K$ is the number of node.

The interaction edge connects the protein-drug domain, similarity edge connects protein-protein domain and drug-drug domain separately, that allows information to flow across the entire DTI graph. What's more, positive edge (e.g. 1) means to pull two nodes closer and negative (e.g. -1) means to push them apart.

\subsubsection{Graph attention network}
Our approach uses Graph Attention Network (GAT)\citep{velikovi2017graph} to adaptively learn weights for each edge and represent each node by message passing.

The input of GAT is a set of node feature $H = (h_1, h_2, ..., h_K), h_i \in R^F$ and DTI adjacent matrix $A$. $H$ contains $H_p = (h_{p_1}, h_{p_2}, ..., h_{p_m})$ and $H_d = (h_{d_1}, h_{d_2}, ..., h_{d_n})$ and $A$ is generated by~\ref{section:graph construction}. 

The output is a new set of node feature, $H^{'} = (h^{'}_1, h^{'}_2, ..., h^{'}_K), h^{'}_i \in R^{F^{'}}$.

In order to transform input features to higher level features, we apply a weight matrix, $W \in R^{F^{'}\times F}$, to each node. 
\begin{equation}
    h^{'}_i = \sigma(W \cdot h_i),  \ h^{'}_i \in R^{F^{'}}
    \label{eq:node feature transform}
\end{equation}

Then we perform a self-attention mechanism on node pair, $a:R^{F^{'}} \times R^{F^{'}} \rightarrow R$, to compute attention weight, which indicates the importance of $n_j$ to $n_i$, $n_j \in  N_i$, where $N_i$ is the neighborhood of $n_i$ and  itself in DTI graph.
\begin{equation}
    w_{ij} = a(h^{'}_i, h^{'}_j), \ w_{ij} \in R^1
    \label{eq:attention1}
\end{equation}
In GAT layer, the attention mechanism $a$ is a single-layer neural network, parametrized by a weight matrix $\widetilde{a} \in R^{2F^{'}}$. Then the LeakyReLU nonlinearity is applied.
\begin{equation}
    w_{ij} = LeakyReLU(\widetilde{a}^{T}[h^{'}_i|| h^{'}_j])
    \label{eq:attention2}
\end{equation}
where $.^T$ is transposition and $||$ is the concatenation operation.

In the general formulation, attention mechanism allows every node to attend on every other node, dropping all structural information. To add graph structure information, we perform mask attention according to the DTI adjacent matrix $A$, which enables only the neighbor nodes to be attended. Then we normalize the attention weight across all choices of $j$ using the softmax function to make it comparable of different nodes.
\begin{equation}
    \alpha_{ij} = Softmax_j(w_{ij}) = \frac{exp(w_{ij})}{\sum_{k \in N_i}exp(w_{ik})}
    \label{eq:softmax}
\end{equation}
where $N_i$ is the set of $h_i$'s neighborhood and itself.

After obtaining the normalized attention score, we use message passing to compute a linear combination of the node features and output the final aggregated features of each node.
\begin{equation}
    h_i^{'} = \sigma (\sum_{j \in N_i}\alpha_{ij}Wh_j)
    \label{eq:message pass}
\end{equation}
where $\sigma$ is a nonlinearity, $H^{'}$ is final aggregated features.

\subsubsection{Final decoder}
The final decoder is a sample neural network, parametrized by a weight matrix $W \in R^{2F^{'}}$. It takes pairs of drug-protein embeddings, generated by GAT layer(e.g. $h_i$ and $h_j$), as input. Then the two node vector do an element-wise multiplication, $p \odot d \rightarrow v, \ v,p,d \in R^F$. Finally, through a layer of neural network $v^F \rightarrow R^1$ and a Sigmoid activate function, produce a probability score indicating whether they interact:
\begin{equation}
    s_{ij} = Sigmoid(ReLU(W(h^{'}_i \odot h^{'}_j)))
    \label{eq:decoder}
\end{equation}

\subsubsection{Loss function}
For training \model, we assign a binary class label ${0,1}$ (interact or not) to each identified drug-protein pair. We assign positive labels to interacted drug-protein pairs and negative labels to non-interacted drug-protein pairs. The output is the interaction probability. 
The main learning objective is to minimize the following binary cross entropy loss (BCELoss) between the target $X = (x_1, x_2, ..., x_N), x_i \in \{0,1\}$ and the output $Y = (y_1, y_2, ..., y_N), y_i \in (0,1)$. The loss can be described as:
\begin{equation}
    l(x,y) =  \{l_1, l_2, ..., l_N\}^T 
    \label{eq:BCELoss1}
\end{equation}
\begin{equation}
    l_n = -w_n[y_n\cdot log \ x_n + (1-y_n)\cdot log \ (1-x_n)]
    \label{eq:BCELoss2}
\end{equation}

\subsubsection{Implementation details}
In this experiment, we use PyTorch \citep{NEURIPS2019_9015} for efficient GPU-based implementation and PyTorch geometric \citep{Fey/Lenssen/2019} for a sparse Graph Attention Layer.

The \model contains the feature (PsePSSM, Fingerprint) encoder, the GAT embedding module and a final interaction decoder. Feature encoder is implemented as two MLP layers for protein and drug respectively. The input dimensions are 220 (PsePSSM) and 881 (fingerprint) and the output dimensions are both 256. The GAT embedding module consists of a Multi-head GAT layer, based on PyTorch geometric \citep{Fey/Lenssen/2019}. The hidden dimension and output dimension both are set to 256 for all datasets. In building the adjacency matrix, the setting of the common neighbor threshold is different in different datasets, it is set to 1 in Nuclear Receptor, 3 in GPCR, Ion Channel and Enzyme dataset. The final interaction decoder includes three MLP layers, which input dimension $2 \times 256 = 512$ and the hidden dimension is also 256. The output is a scalar, then be computed by a sigmoid function to get the final interaction probability.

In our paper, the \model are trained end-to-end using gradient descent based on Adam optimizer \citep{kingma2014adam}. We perform batch gradient descent using the full dataset for every training iteration, which is a viable option as long as datasets fit in memory. Base on the implementation of PyTorch Geometric \citep{Fey/Lenssen/2019}, we use a sparse representation for A. The memory requirement is $\mathcal{O} (|\varepsilon|)$, which is linear in the number of edges. The learning rate $\alpha$ is 0.0005 and the weight decay is $5e-4$. Stochasticity in the training process is by dropout \citep{JMLR:v15:srivastava14a}, which is set to 0.3. The model is trained until converging for each fold of the cross-validation,
for which we set the epoch number to be 4000, 8000, 10000, 15000 for Nuclear Receptor, GPCR, Ion Channel and Enzyme, respectively.

\section{Results}\label{sec:result}
\begin{table}[]
\caption{Number of drugs, target proteins, and their interactions for 5 datasets used for cross-validation.}
\centering
\begin{tabular}{|c|c|c|c|c|c|}
\hline
             & Enzyme & IC   & GPCR & NR & Drugbank\_approved      \\ \hline
Drugs        & 445    & 210  & 223  & 54 & 1555                    \\ \hline
Targets      & 663    & 204  & 95   & 26 & 1591                    \\ \hline
Interactions & 2925   & 1476 & 635  & 90 & 5831                    \\ \hline
\end{tabular}
\label{tab:data}
\end{table}


\begin{figure*}
        \centering
        \begin{subfigure}[b]{0.495\textwidth}
            \centering
            \includegraphics[width=\textwidth]{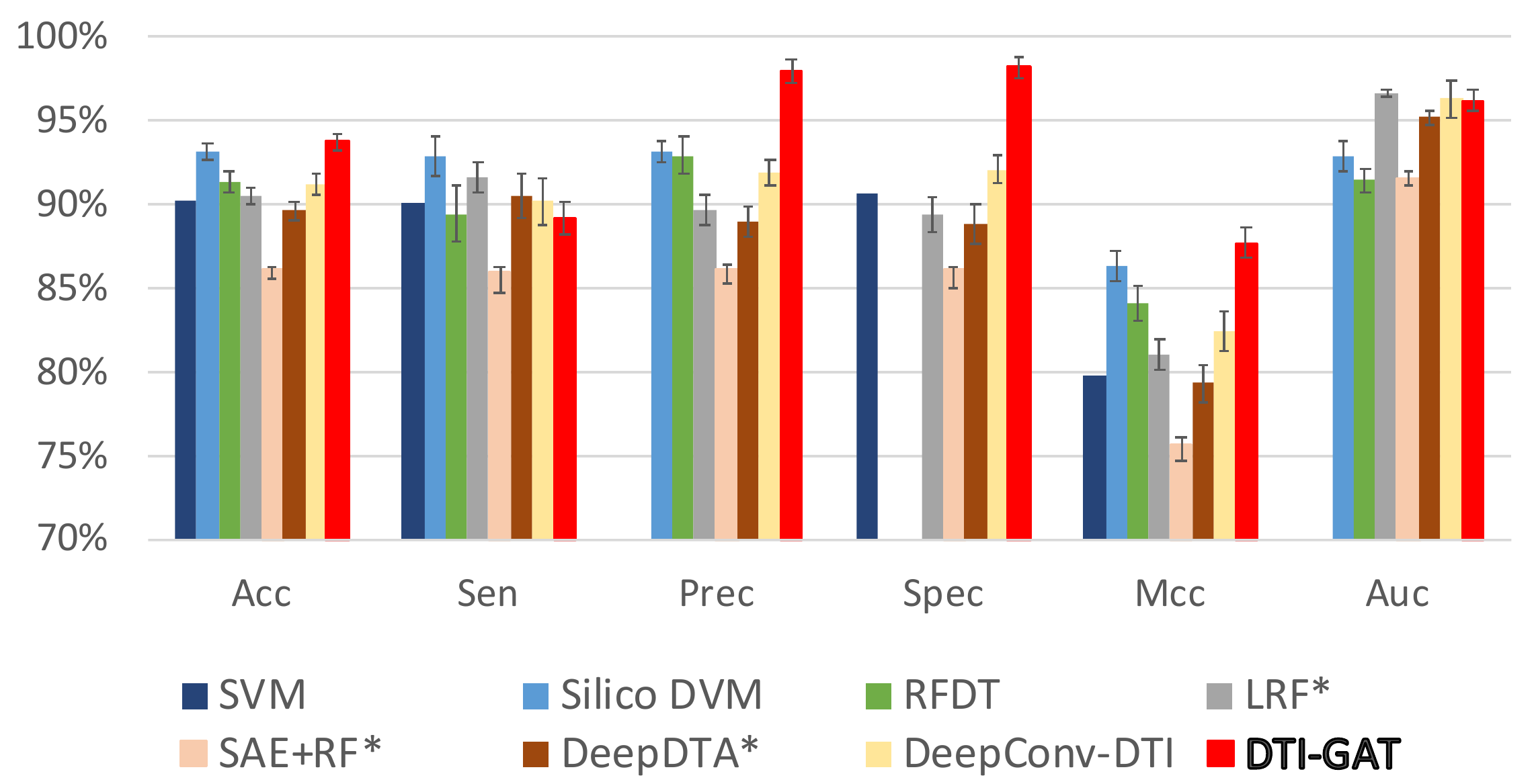}
            \caption[Enzyme]%
            {{\small Enzyme}}    
            \label{fig:mean and std of net14}
        \end{subfigure}
        \hfill
        \begin{subfigure}[b]{0.495\textwidth}  
            \centering 
            \includegraphics[width=\textwidth]{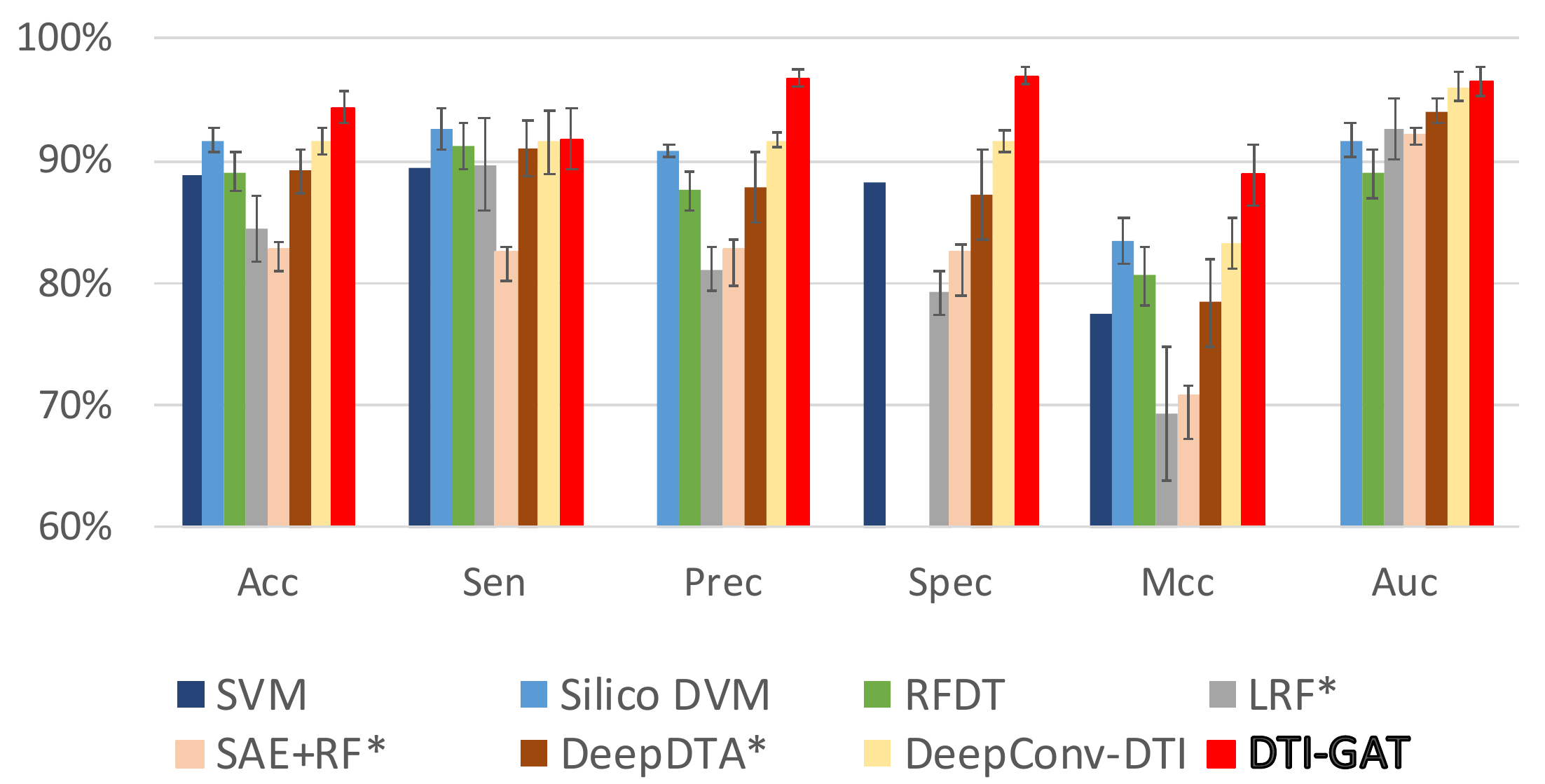}
            \caption[]%
            {{\small Ion Channel}}    
            \label{fig:mean and std of net24}
        \end{subfigure}
        \vskip\baselineskip
        \begin{subfigure}[b]{0.485\textwidth}   
            \centering 
            \includegraphics[width=\textwidth]{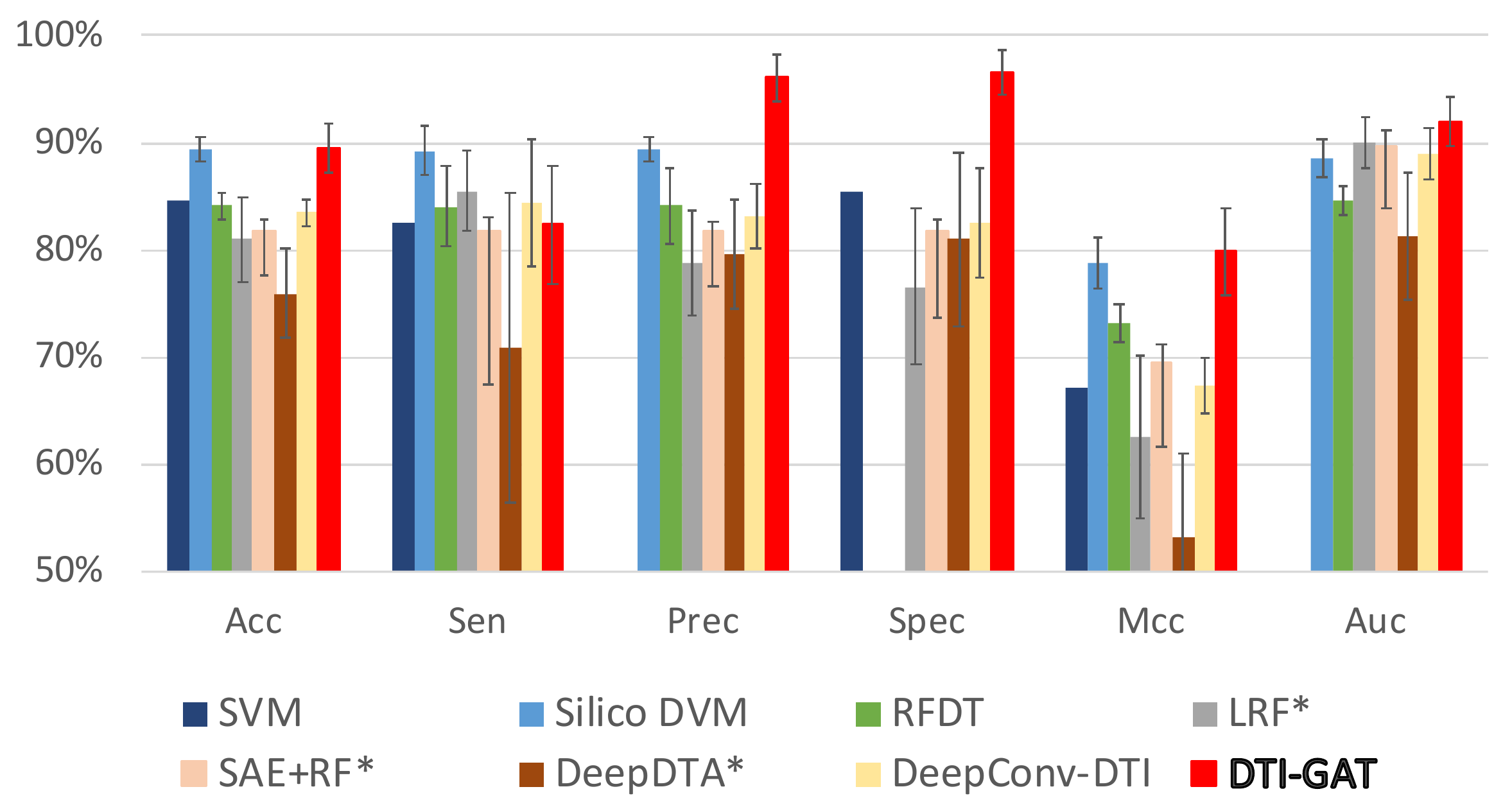}
            \caption[]%
            {{\small GPCR}}    
            \label{fig:mean and std of net34}
        \end{subfigure}
        \quad
        \begin{subfigure}[b]{0.485\textwidth}   
            \centering 
            \includegraphics[width=\textwidth]{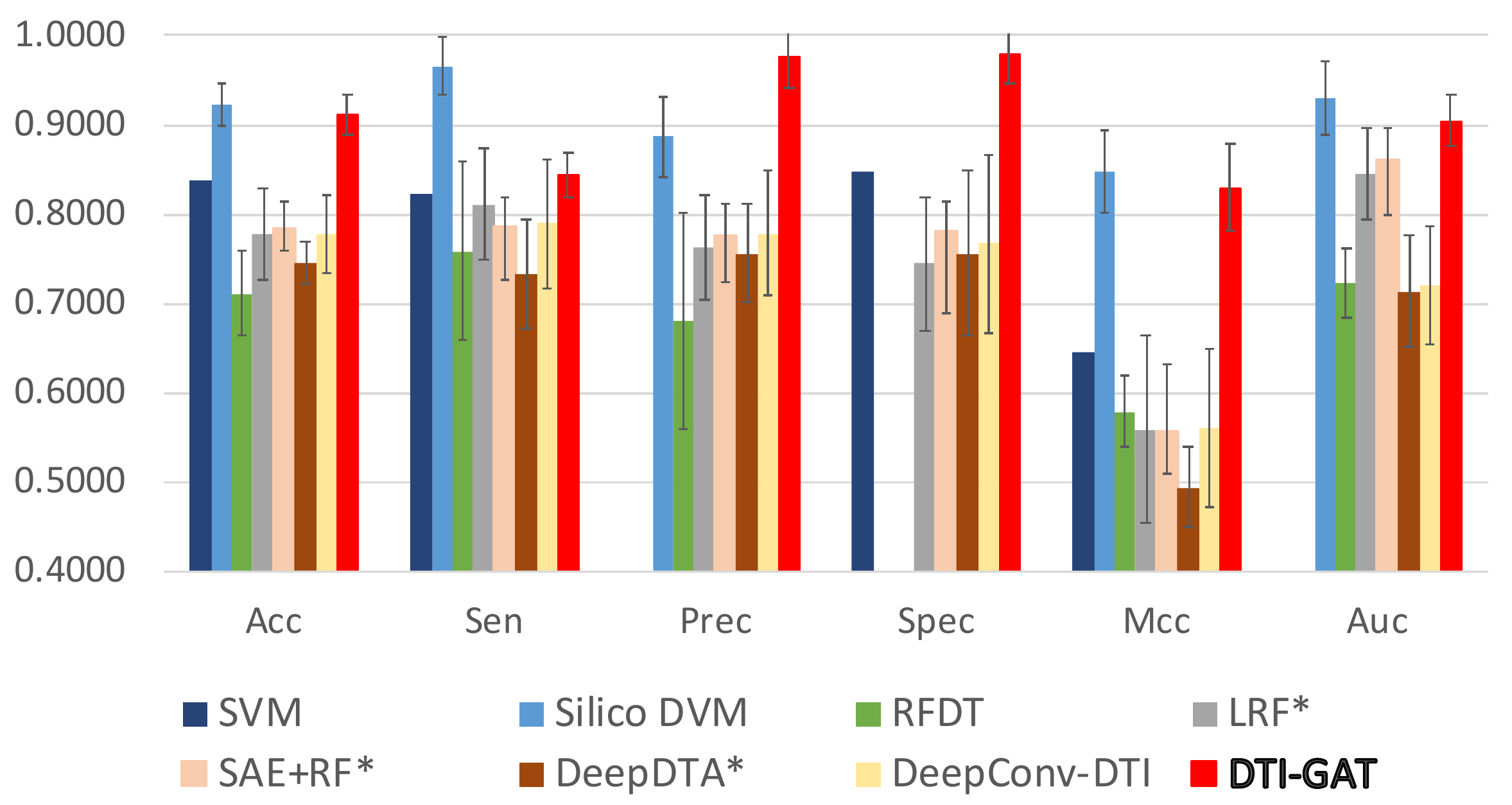}
            \caption[]%
            {{\small Nuclear Receptor}}    
            \label{fig:mean and std of net44}
        \end{subfigure}
        \caption[]
        {Evaluation of DTI prediction on the four datasets based on 5-fold cross-validation under 6 evaluation metrics. We report the mean and standard deviation for the test sets for all methods.} 
        \label{fig:main_eval}
    \end{figure*}
    
\begin{table}[]
\caption{Evaluation of DTI prediction on the drugbank\_approved dataset based on 5-fold cross-validation under 6 evaluation metrics. We report the mean and standard deviation for the test sets.}
\centering
\begin{tabular}{|l|l|l|l|l|l|l|l|}
\hline
 & & Acc          & Sen  & Prec    & Spec    & Mcc     & Auc\\\hline
SAE+RF*      & mean & 82.60 & 82.75 & 82.39 & 82.54 & 66.05 & 88.31 \\
             & std  & 1.24  & 1.35  & 1.16  & 1.07  & 1.29  & 0.31\\\hline 
DeepDTA*     & mean & 82.13 & \textbf{83.61} & 81.29 & 80.65 & 64.34 & 87.36 \\
             & std  & 0.96  & 1.76  & 2.15  & 2.94  & 1.92  & 0.77  \\\hline
DeepConv-DTI & mean & 84.54 & 83.44 & 85.31 & 85.65 & 69.02 & 91.73 \\
             & std  & 0.65  & 0.72  & 0.83  & 0.98  & 1.21  & 0.34  \\\hline
LRF*         & mean & 86.11 & 83.45 & 88.18 & 88.78 & 72.38 & 92.20 \\
             & std  & 0.45  & 2.40  & 1.13  & 1.56  & 0.77  & 0.75  \\\hline    
\model      & mean & \textbf{87.67} & 81.12 & \textbf{93.34} & \textbf{94.21} & \textbf{76.00} & \textbf{92.30} \\
             & std  & 0.95  & 2.24  & 0.60  & 0.65  & 1.69  & 0.83  \\\hline             

\end{tabular}

\label{tab:drugbank_eval}
\end{table}    

\begin{figure}[!t]
    \centering
        \includegraphics[width=\linewidth]{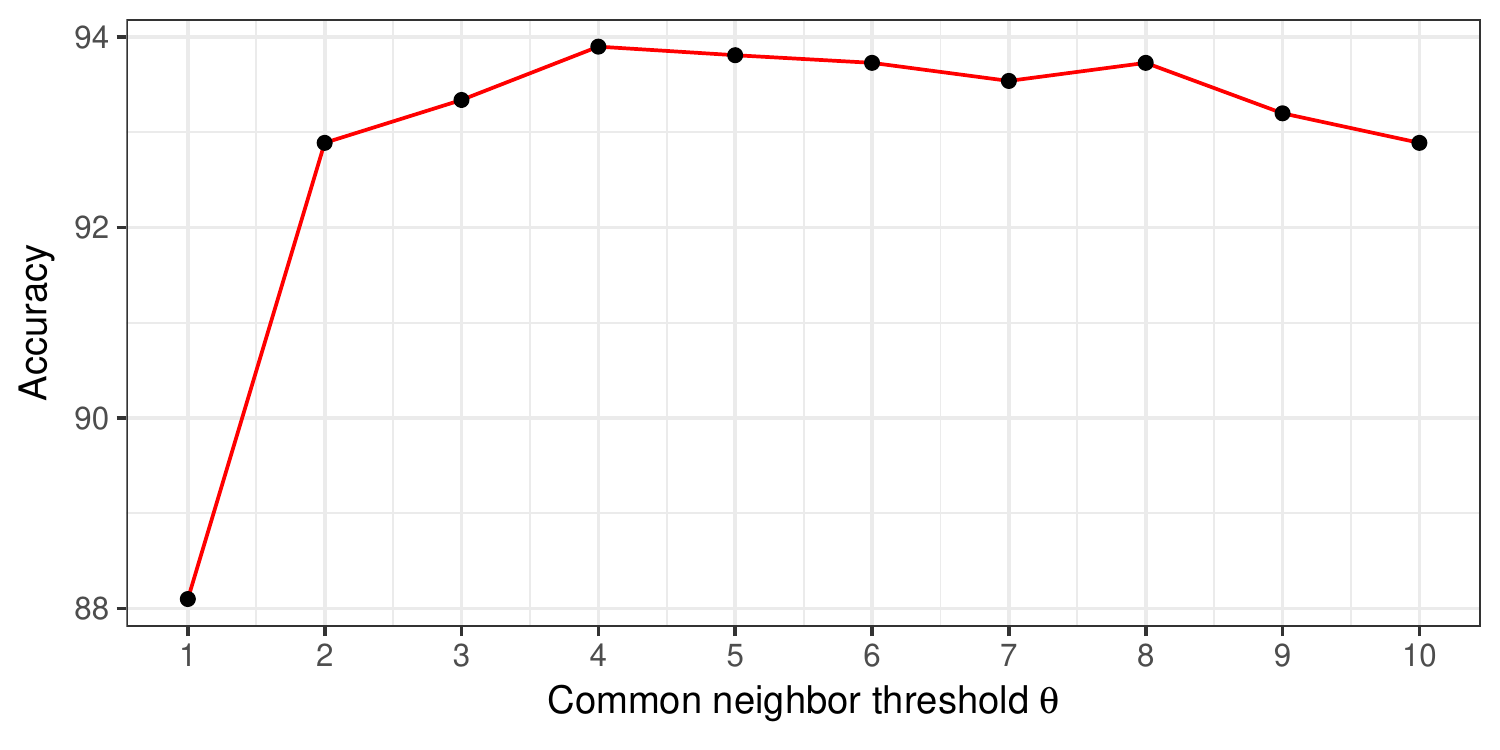}
        \label{fig:hyper1}
    
    
    \caption{Performance evaluation on using different hyperparameters for PDI prediction on the Enzyme dataset. The accuracy is reported.}\label{fig:hyper}
    
\end{figure}

\begin{table}[]
\caption{Evaluation of DTI predictions by different methods on the independent study set.}
\centering
\begin{tabular}{|l|l|l|l|l|l|l|}
\hline
             & Acc   & Sen   & Prec  & Spec  & Mcc   & Auc   \\ \hline
SAE-RF*       & 64.20 & 55.30 & 69.30 & 77.30 & 32.10 & 71.20 \\ \hline
DeepDTA*      & 64.82 & 67.33 & 59.80 & 62.75 & 29.95 & 73.04 \\ \hline
DeepConv-DTI & 68.60 & 58.50 & 73.20 & 78.60 & 37.90 & 74.30 \\ \hline
LRF*          & 71.38 & 45.89 & 73.19 & \textbf{92.37} & 44.02 & 75.70 \\ \hline
\model      & \textbf{72.90} & \textbf{67.36} & \textbf{75.54} & 78.39 & \textbf{46.04} & \textbf{78.42} \\ \hline

\end{tabular}

\label{tab:indep_eval}
\end{table}    
    
\begin{figure*}
        \centering
        \begin{subfigure}[b]{0.495\textwidth}
            \centering
            \includegraphics[width=\textwidth]{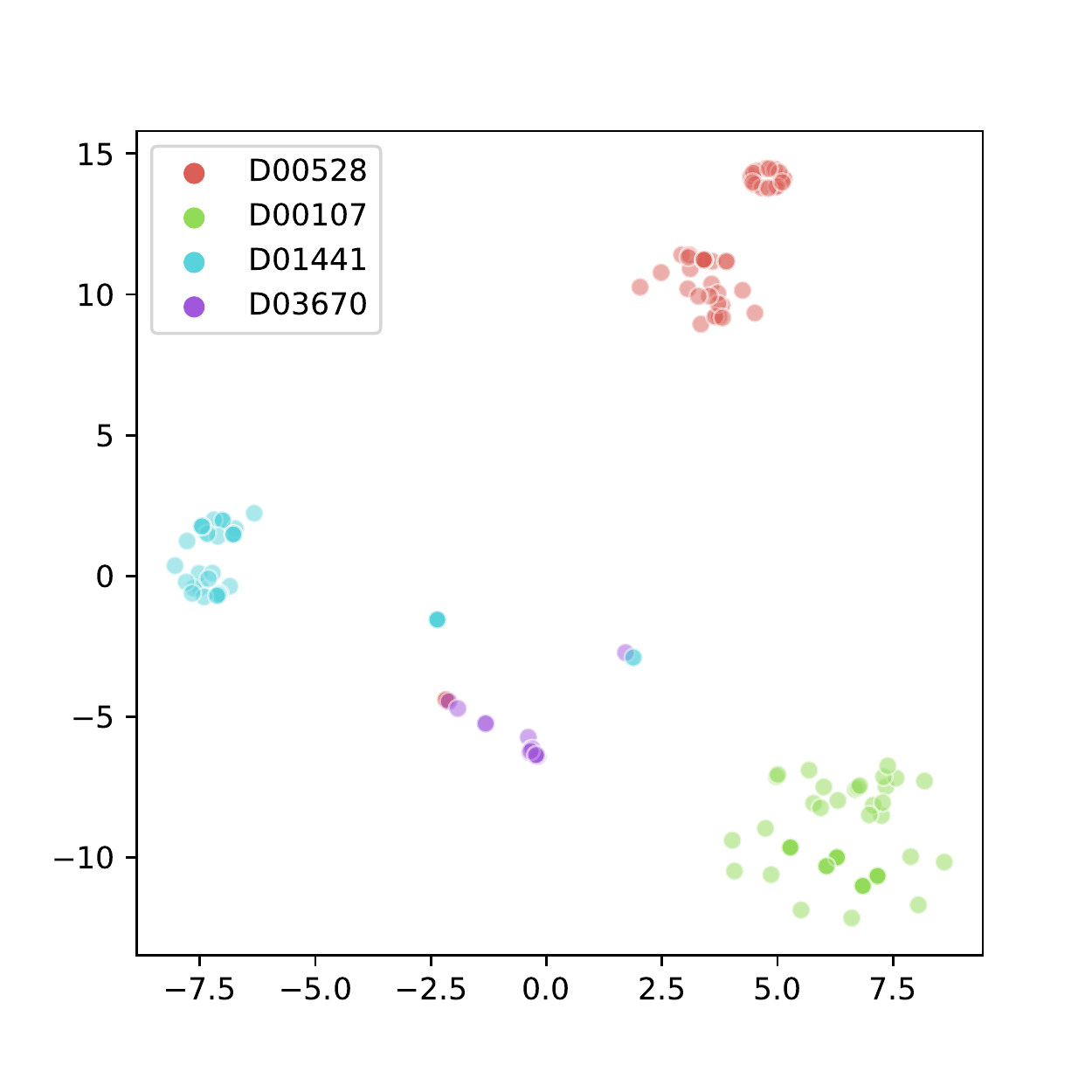}
            \caption[Enzyme]%
            {{Protein nodes grouped by top connected drugs}}    
            \label{fig:tsne_protein}
        \end{subfigure}
        \hfill
        \begin{subfigure}[b]{0.495\textwidth}  
            \centering 
            \includegraphics[width=\textwidth]{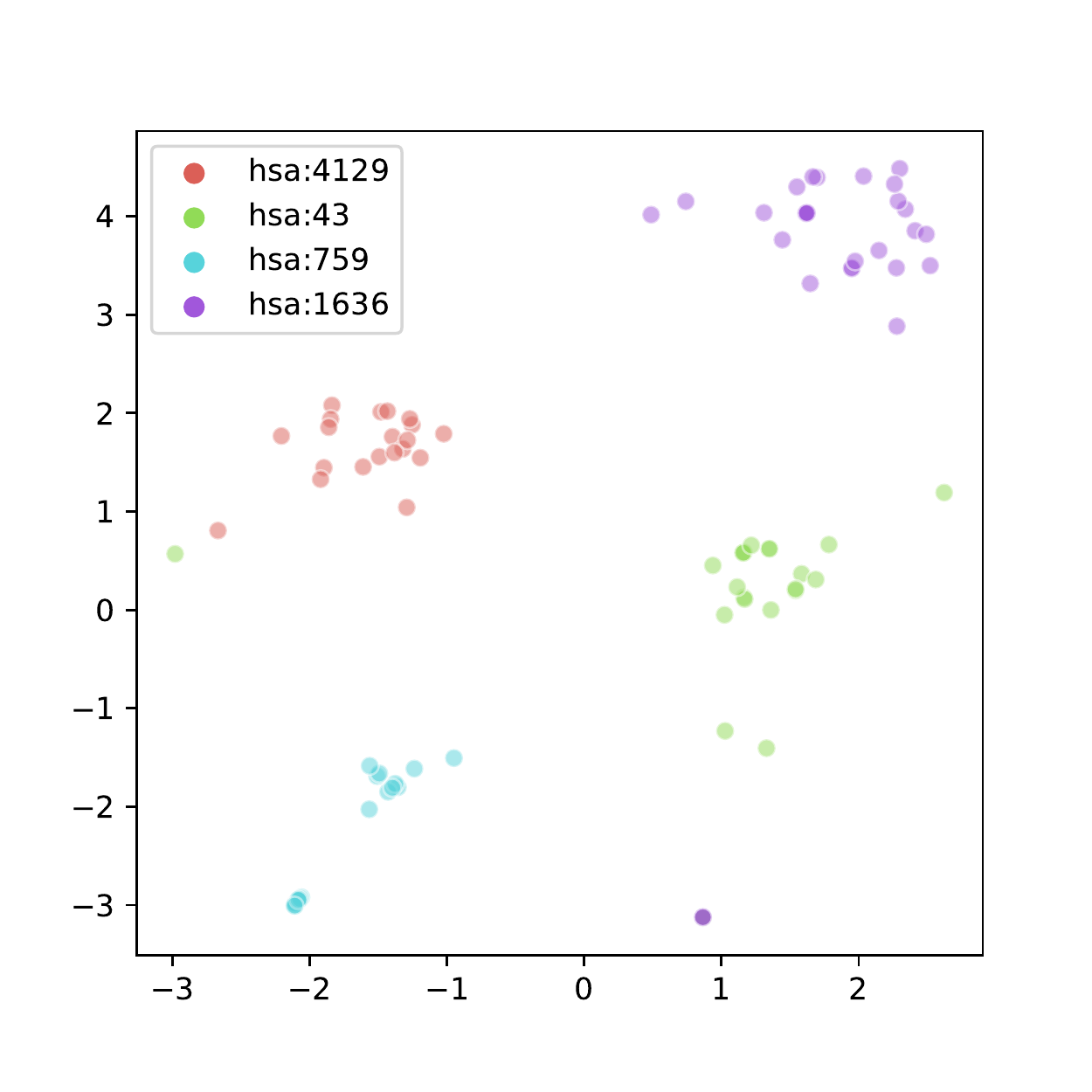}
            \caption[]%
            {{Drug nodes grouped by top connected proteins}}    
            \label{fig:tsne_drug}
        \end{subfigure}
        \caption[t-SNE visualization of the node embeddings generated by \model on the Enzyme dataset.]
        {t-SNE visualization of the protein (a) and drug (b) nodes embeddings generated by \model on the Enzyme dataset.} 
        \label{fig:tsne_all}
    \end{figure*}    

We present the experimental evaluation of the proposed framework on the binary DTI prediction task. The experiments are conducted on the following datasets.

\subsection{Datasets}
We use the interaction data between drugs and target proteins collected by \cite{yamanishi2008prediction}, available at http://web.kuicr.kyoto-u.ac.jp/supp/yoshi/drugtarget. The dataset is collected from various databases like SuperTarget \citep{gunther2007supertarget}, DrugBank \citep{wishart2008drugbank}, KEGG BRITE \citep{kanehisa2006genomics}, and BRENDA \citep{schomburg2004brenda}. 
This dataset includes four main subsets: enzymes, ion channels (IC), G-protein-coupled receptors (GPCR) and nuclear receptors (NR). The statics of the interaction are shown in Table~\ref{tab:data}. We obtained the protein sequence information from UniProt \citep{uniprot2019uniprot}, and the drug fingerprint data from PubChem \citep{kim2019pubchem}. Note that there are a few proteins and drugs that have been removed in the newer version of databases, so we did not include them in our training. 

In addition to the above four benchmark datasets, we also use a dataset called drugbank\_approved \citep{ba2016daspfind}, which contains all FDA-approved drugs and their corresponding protein targets in the DrugBank database \citep{wishart2008drugbank}. After removing the non-existing proteins and drugs, the processed dataset contains 1555 drugs and 1591 targets. 



\subsection{Evaluation protocol}
Following the settings in previous works \citep{wang2018rfdt, wang2018computational, shi2019predicting}, we conduct 5-fold cross-validation (CV) on the five datasets of the enzyme, ion channel, GPCR, nuclear receptor and the drugbank\_approved.
Under the 5-fold CV setting, the data is equally divided into 5 non-overlapping subsets, and each subset has a chance to train and to test the model so as to ensure an unbiased evaluation. We repeat the prediction model with 10 trials and record the mean and standard deviation of results.
We aggregate fix metrics on the test cases of each fold, i.e. the overall \emph{accuracy (ACC)}, \emph{precision (PR)}, \emph{sensitivity (SE)}, \emph{specificity (SP)}, \emph{Matthews correlation coefficient (MCC)} and \emph{Area under the curve (AUC)}.
All these metrics are preferred to be higher to indicate better performance. 

\subsection{Baseline methods}
We compare \model with the following two groups of baselines:
\begin{enumerate}
    \item Classical statistical learning models:
    SVM\citep{cao2012large}, Silico DVM\citep{li2017silico}, and LRF\citep{shi2019predicting}.
    \item Deep learning models:
    SAE+RF \citep{wang2018computational}, DeepDTA \cite{ozturk2018deepdta}, and DeepConv-DTI \cite{lee2019deepconv}. 
\end{enumerate}

Note that LRF has been modified to perform data augmentation on the training set only, to ensure validation and training samples are fully separated. 
Also, note that DeepDTA is designed for Drug-Target Affinity Regression instead. We adapt it by using a threshold of 0.5 to do a binary classification in order to obtain the interaction.

\subsection{Experimental results}
\subsubsection{Comparison with other methods}
For the prediction of drug-target interactions, many prediction methods have been proposed. Figure~\ref{fig:main_eval} details the comparison between \model with other baseline methods on enzymes, ion channels, GPCRs, nuclear receptors and Table~\ref{tab:drugbank_eval} shows the results on the larger drugbank\_approved datasets.

As shown in Figure~\ref{fig:main_eval}, in general, the statistical baselines perform better than the deep learning methods. This shows that deep learning methods are not able to obtain enough information to generalize.
The best performing model among them is Silico DVM, a statistical learning model that trains a discriminative vector machine classifier with a local binary pattern generated from PSSM.
Our model \model can generalize using additional information from the graph, thus achieving better performance than the baseline models. \model outperforms Silico DVM on Enzyme, Ion channel and GPCR by 0.014, 0.054, 0.011 on MCC, but Silico DVM outperforms \model on nuclear receptor by 0.018 on MCC. This attribute to the fact that \model is able to extract more generalizable information from the graph of interaction, especially on a large, dense graph. 

Specifically, we can see that for the dataset of the enzyme, the accuracy rate of \model reaches 93.75\%, which is higher than other prediction methods. In addition, the AUC of the prediction model reaches 96.32\%, comparable with the best performing model LRF of 96.64\%. 
In the ion channel dataset, the accuracy of prediction in this paper is 94.38\%, which is also higher than other prediction methods. Our AUC of 96.51\%is the highest among other baselines. Similarly in the GPCR dataset, \model outperforms all baselines in accuracy, precision, specificity, MCC and AUC. As for the nuclear receptor dataset, our prediction method accuracy rate is 91.11\% and our AUC is 90.49\%, slightly lower than the best performing Silico DVM. This is mainly due to the fact that the nuclear receptor dataset only contains a very limited number of samples for training and all the deep learning based methods suffers from it. However, we still drastically outperform other neural network based methods including SAE+RF, DeepDTA and DeepConv-DTI. 

The results on the drugbank\_approved dataset are shown in Table~\ref{tab:drugbank_eval}. With this larger dataset, we focus on comparing our model with four other methods including LRF, DeepConv-DTI, DeepDTA and SAE+RF. 
We can see from the table that the performances for all the methods drop by 5\% to 10\% comparing to the previous cross-validation results. This is because the larger drugbank\_approved dataset contains more complicated and new protein-drug interactions, which make the learning process harder.
However, the accuracy rate of \model reaches 87.67\% and the AUC of \model reaches 92.30\%, which are still higher than all the baseline methods and relatively robust in identifying new interactions.  

\subsubsection{Independent Study}

To show the generalizability of \model, we evaluate the performance of different methods on an independent dataset. 
We construct the independent dataset using the five datasets we discussed in Table~\ref{tab:data}. This independent study dataset consists of a training set of enzyme, IC, GPCR and NR. The testing set comprises of the drugbank\_approve dataset, with the protein-drug pairs that appear in the training set removed. Since the proteins and drugs have different names in different sources, we perform the matching and removal by directly comparing the PsePSSM and the fingerprints. 
After removing the similar sequences, we obtain the independent dataset with 1883 drugs and 2048 proteins. Among them, 5113 interactions are used for training and the rest 5315 interactions are used for independent testing.
Using this dataset, we can examine the ability of \model to generalize the prediction to a larger protein network from a subset of networks. 

We compare \model with four baselines including SAE+RF, DeepDTA, DeepConv-DTI, and LRF.
As shown in Table~\ref{tab:indep_eval}, \model outperforms other methods by achieving the highest accuracy of 72.9\%, sensitivity of 67.36\%, MCC of 46.04\% and AUC of 78.42\%. Notice that LRF achieves high specificity with the cost of getting very low sensitivity, meaning that there are many false negatives (predicting many interactions to be non-interactions), which is not optimal for the DTI prediction task. 
The comparative results of \model on the independent set demonstrate that \model can generalize better than other methods to drugs and targets that share low sequence similarity with the training set.

\subsubsection{Parameter analysis}
In our work, the main hyperparameter is the common neighbor threshold, $\theta$. It is the key factor of similarity edge, $e_s$, for which links the protein and drug domain. 

To further discuss the influence of common neighbors, we choose the Enzyme dataset, which has the most nodes. The result of different settings is illustrated by Figure~\ref{fig:hyper}. $\theta = 1$ means once two nodes have common neighbors, they will be linked. In our observation, there are some clustered graphs, many different drug nodes are linked to the same protein node. Once $\theta = 1$, these drug nodes will be joined into a fully connected graph, the DTI graph structure and similarity information will be destroyed. That's why the performance of $\theta = 1$ is lower than the others.  As the $\theta$ increases, the DTI graph includes more and more useful similarity edges, leads to higher accuracy. What's more, if we delete all the $e_s$, means $\theta$ is infinite, no similarity edge, the score is only 87.01, much lower than others, that proves it is very important.  

\subsubsection{t-SNE visualization of the node embeddings by \model}
The effectiveness of the learned feature representations for each node may also be investigated qualitatively. For this purpose, we provide a visualization of the t-SNE \citep{maaten2008visualizing} transformed feature representations extracted by the output layer of the GAT model on the Enzyme dataset. 
The t-SNE visualizations are generated for both the view of protein nodes and the view of drug nodes as shown in Figure~\ref{fig:tsne_protein} and Figure~\ref{fig:tsne_drug}, respectively. 

To show the 2D-projection of protein features, we choose the top 4 drugs as labels ordered by their number of connections to all proteins. We then filter out proteins that are not connecting to any of the selected drugs or ambiguously connecting to multiple selected drugs. The feature vectors of the rest proteins outputted by GAT are then transformed into 2D space by t-SNE.
Similarly, we choose the top 4 proteins as labels and conduct the same filtering process on the drugs to visualize the drug node features. 

As shown in Figure~\ref{fig:tsne_protein}, each dot represents a protein and each color represents the drug that the protein is interacting with. Most proteins except the two blue ones are forming discernible clusters. Note that these clusters correspond to the labels of the common interacting drugs. Similarly in Figure~\ref{fig:tsne_drug}, each dot represent a drug and each color represent the protein that the drug is interacting with. We can also see four clusters are formed based on the learned features of each drug. These two different views of visualizations in 2D space by t-SNE demonstrate that our learned attended vector can capture the important interactions while preserving the properties of each node. 

To demonstrate the biological interpretability of the learned features, in addition to the internal distance within each cluster, we also examine the distance among different clusters. The distance also correlates with the pathway of the labeled proteins. For example, in Figure~\ref{fig:tsne_drug}, the blue hsa:759 cluster has only metabolic pathways, and the green hsa:43 and red hsa:4129 clusters both participate in metabolism and other functions. These 3 clusters share the metabolic pathway.
The purple one, hsa:1636 instead has pathways in the Renin system. Therefore, hsa:759 is closer to hsa:43 and hsa:4129 comparing to the distance between hsa:759 and hsa:1636.

\subsubsection{Attention weight analysis}



\begin{figure}[!t]
    \centering
    \includegraphics[width=\columnwidth]{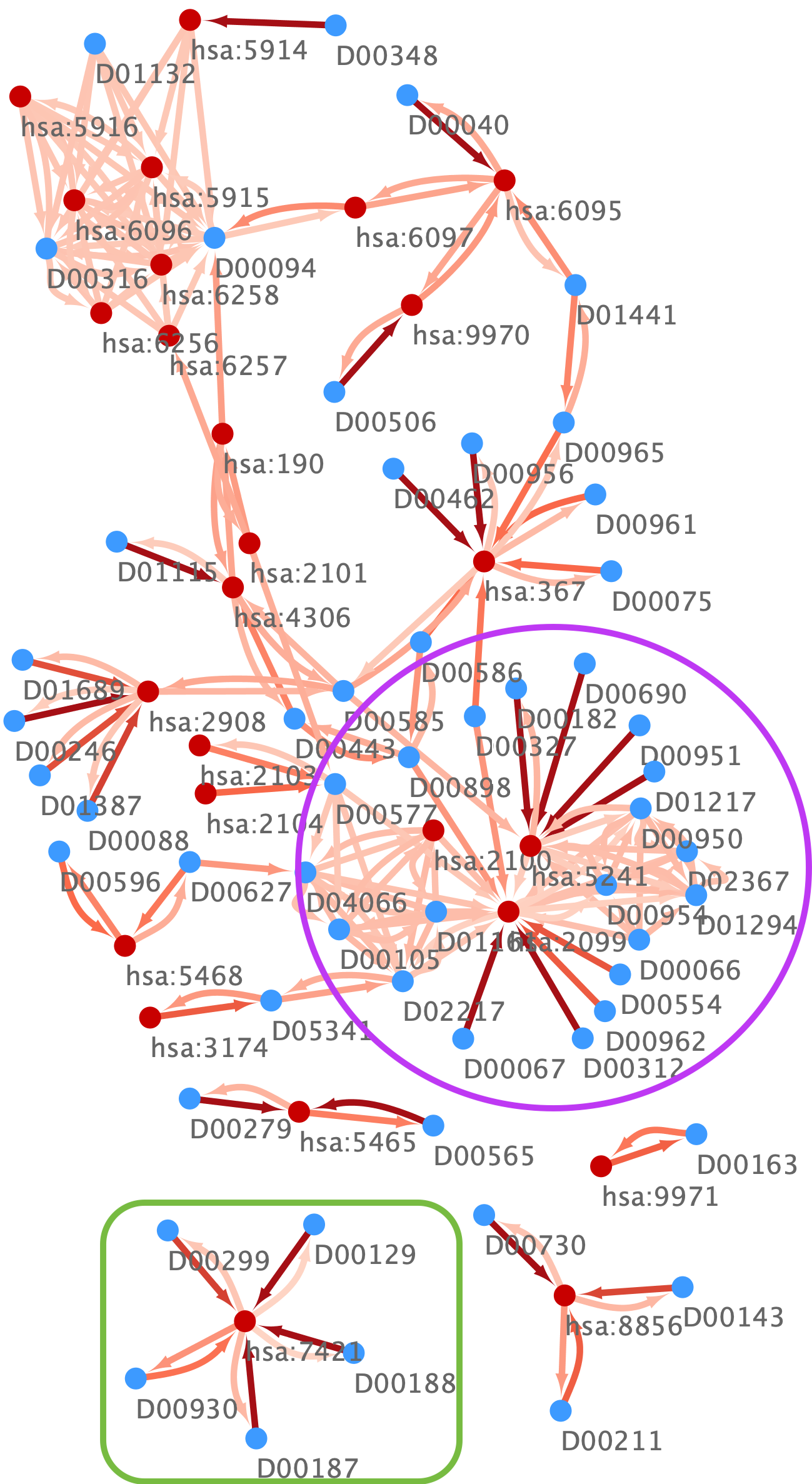}
    
    \caption{Attention graph of \model on NUC.}
    \label{fig:att_val}
\end{figure}

To further illustrate the effectiveness of the graph attention approach, we also compute an attention graph from the training, shown in Figure~\ref{fig:att_val}. We generate this figure using the attention value of the edges trained from the Nuclear Receptor dataset. From this graph, the proteins and drugs are able to form groups based on their similarities in the functions. We focus on analyzing two main groups. 

First, we analyze the large group around proteins hsa:2099  (estrogen receptor alpha), hsa:2100 (estrogen receptor beta), and hsa:5241 (progesterone receptor), shown in the purple circle in the figure. The functions of these proteins are very similar. Therefore they are shown as connected and interact with many drugs. The drugs to the right of these proteins consist of a few progesterone receptor agonists (D00182, D00951, D01294, D00066, D00950, D00954, D01217, D02367). For some of the drugs, they only act as progesterone receptors, such as D00951 and D01217, and therefore have a large weight connecting to the progesterone receptor. On the other hand, drugs like Progesterone (D00066) also activate the estrogen pathway and therefore are connected to the estrogen receptor pathway. 
Because of there similarity, these drugs also give attention to other drugs with a similar function, lowering the individual attention weight in the cluster. 
On the smaller cluster on the left of the estrogen receptors, we can also see a similar cluster of estrogen receptor agonists (D00554, D00105, D00067, D00312, D00577, D00898) and estrogen receptor antagonists (D01161). These proteins do not activate progesterone receptors, with a few of them only activate the alpha receptor (D00962, D02217). 

In addition, we analyze the smaller cluster on the top right of the graph, shown in the green box. This cluster is disjoint from the rest of the graph, centered around Vitamin D Receptor (hsa:7421). There are 5 drugs around this receptor, which are the following: 

\begin{itemize}
    \item Vitamin D2 (Ergocalciferol, D00187)
    \item activated Vitamin D2 (Paricalcitol, D00930)
    \item Vitamin D3 (Cholecalciferol, D00188)
    \item activated Vitamin D3 (Calcitriol, D00129)
    \item a synthetic vitamin D analog (Dihydrotachysterol, D00299)
\end{itemize}

For these drugs, we can see that the receptor has placed more attention weights on the activated versions of the vitamin D (D00930 and D00129). And for these two drug, there are less attention on the receptor. This is because they pay more attention on itself, lowering their attention weight to the receptor. However, our graph is not able to predict the connection between the vitamins and their activated version due to insufficient information. 

We can see that \model is able to learn the importance of each interaction from the interaction graph of the dataset. This allows the model to effectively predict the interaction between the drug and the protein. 

\section{Conclusion}
In this paper, we introduce a novel and comprehensive learning framework to predict the drug-target interactions. 
Our proposed framework, \model, operates on the graph-structured data with the attention mechanism based on the deep neural network architecture. 
We provide an approach to transform the feature representations of proteins and drugs into a protein-drug interaction graph. We also emphasize the importance of adding the drug-drug and protein-protein similarities to the graph. 
With the attention mechanism, \model can automatically extract the important high-level relationships by assigning different weights to each edge.
Extensive experiments of both cross-validation and independent tests conducted on the five datasets demonstrate the promising performance of \model. 
Moreover, we provide case studies to show the interpretability of \model. 
As a future direction, we plan to explore other external knowledge including gene ontology and PPI networks, which can be integrated into \model. To further improve the performance, we seek to incorporate the knowledge graph representation learning framework \citep{hao2019universal}.





\bibliographystyle{natbib}
\bibliography{reference}

\end{document}